\documentclass[aps,prb,epsfigm,twocolumn,showpacs]{revtex4}
\usepackage{graphicx}
\usepackage{amsmath}
\usepackage{amsfonts}

\begin{document}

\title{The Role of Alternance Symmetry in Magnetoconductance}

\author{Josep Planelles, Juan I. Climente}
\affiliation{Departament de Qu\'{\i}mica F\'{\i}sica i Anal\'{\i}tica,
Universitat Jaume I, E-12080, Castell\'o, Spain}
\email{josep.planelles@uji.es}
\homepage{http://quimicaquantica.uji.es/}
\date{\today}

\begin{abstract}
We show that the direction of coherent electron transport across a cyclic system of quantum dots or a cyclic molecule
can be modulated by an external magnetic field if the cycle has an odd number of hopping sites, but the transport
becomes completely symmetric if the number is even.
These contrasting behaviors, which remain in the case of interacting electrons,
are a consequence of the absence or presence of alternance symmetry in the system.
These findings are relevant for the design of nanocircuits based on coupled quantum dots or molecular junctions.
\end{abstract}


\maketitle

\section{Introduction}

Electron transport across cyclic quantum dot molecules is a subject of active research in condensed matter physics,
where 'artificial molecules' made of tunnel-coupled quantum dots with looped structure can be fabricated.\cite{
GaudreauPRB,TarankoJAP} The interest of these systems follows from the quantum-mechanical interference of 
electronic paths, which gives rise to a variety of physical phenomena.
In recent years, much attention has been paid to triple-dot structures, owing to their potential for
the development of spin qubits.\cite{GaudreauNP} Four-dot structures have also been proposed
for spin qubits because they constitute the smallest loop enabling decoherence-free subspaces.\cite{LidarPRL,ScarolaPRL}

In parallel, there is also a current endeavor to develop molecular wires and molecular 
electronics based on single-molecule juctions.\cite{NitzanSCI,TaoNNT,SongAM}
 Also here there is interest in conjugated cyclic molecules, 
as their multiply-connected topology is expected to enable conductance modulation by means of the
quantum interference processes.\cite{LiuJACS,HodJACS,HodJPCM,HansenJCP,MarkussenNL}

Clear manifestations of the topology of cyclic systems are expected in their magnetoconductance,
as the magnetic flux enclosed by the delocalized electron translates into an Aharonov-Bohm (AB) phase\cite{AharonovPR}
whose consequences on the electronic transport have been discussed in the literature, 
be it for aromatic hydrocarbons\cite{HodJACS,HodJPCM,MardaaniJMMM}, for coupled quantum dots\cite{GaudreauPRB,DelgadoPRL,JiangJPCM,EmaryPRB,KuzmenkoPRL},
or for ring-shaped nanostructures.\cite{WebbPRLTimpPRL,ChwiejJPCMIjasPRB,SzafranEPL} 

While the role of spatial and spin symmetries in the transport and magnetotransport of 
cyclic systems has been thoroughly investigated,\cite{DelgadoPRL,KuzmenkoPRL,MarkussenNL,TaniguchiJACS,LortscherCPC}
much less is known about the role of the so-called alternance symmetry.
From the early times of quantum chemistry, alternance symmetry has proved very useful in 
predicting the properties of conjugated hydrocarbons. 
This topology-related symmetry classifies the conjugated hydrocarbons into two non-overlapping groups:
alternant or non-alternant.\cite{Orchin_book}
The idea is to divide all carbons in a molecule into two sets, one marked with stars ($\star$) and the 
other with circles ($\circ$). The system is alternant if it is possible to place
stars (circles) on alternating carbons, with no two stars (circles) adjacent.
Thus, hydrocarbons with odd-membered cycles are non-alternant, 
while linear structures and even-membered cycles are alternant. 
The bipartite nature of alternant systems gives rise to a particle-hole symmetry in
the single particle energy spectrum which is missing in non-alternant systems. As a result, molecules
of the each kind share features of the electronic structure whose impact on the 
reactivity and spectroscopy have been long recognized.\cite{PariserJCP,MichlJCP,CallisJPC,Jones_book}

In this work, we study single-electron transport across planar cyclic systems subject to an external axial magnetic field. 
The goal is to gain understanding on the role of alternance symmetry on such transport.
To this end, we model both alternant and non-alternant systems using a Hubbard Hamiltonian.
The cyclic systems are coupled to one input and two output channels, forming a three-terminal device.
When the system is non-alternant, the magnetic field is shown to modulate the direction
of electron transport, favoring the transfer probability through one of the output channels.
When the system is alternant, by contrast, the transport is completely symmetric through
both output channels. An interpretation is provided for these results by analyzing the
alternance symmetry in the Hamiltonian.
The influence of excess interacting electrons in the molecule is also investigated.
It is found that the role of alternance symmetry holds in the few-electron case.
 
The paper is organized as follows. In Section \ref{s:theo} we introduce the theoretical
model. In Section \ref{s:res} we describe the numerical simulations of electron magnetotransport
in both alternant and non-alternant molecules. In Section \ref{s:disc} we discuss how these results
may relate to either molecular devices using aromatic molecules or solid state quantum dot
circuits. Finally, in Section \ref{s:conc} we give conclusions.

\section{Theoretical Considerations}
\label{s:theo}

\subsection{Hamiltonian and time evolution}

In order to describe our system of interacting fermions we use a Hubbard Hamiltonian.
The Hamiltonian, built on a nearest-neighbor tight-binding (TB) formalism and including a perpendicular magnetic field, reads:
\begin{equation}
\label{eq1}
H= \sum_i \epsilon_i  c_i^{\dag} c_i  +  \sum_{<ij>} [t^0_{ij} \; e^{i\theta_{ij}} c_i^{\dag} c_j + H.c.] + U \sum_i (n_i^2- n_i)/2
\end{equation}
\noindent where $c_i^{\dag}$ ($c_i$) denotes the creation (annihilation) operator of the site $i$, $n_i$ is the number operator and
$\epsilon_i$ is the on-site potential at the $i$th site, that we set equal to zero unless otherwise indicated.
$t^0_{ij}$ is the zero-field tunneling parameter between the nearest-neighbor sites $i$ and $j$, 
$\theta_{ij}$ is the Peierls phase given by $\theta_{ij}=(2\pi/\Phi_0) \int_i^j {\mathbf A} \cdot d{\mathbf l}$, 
with $\Phi_0=hc/e$ being the magnetic flux quantum and ${\mathbf A}$ the vector potential.\cite{GrafPRB} 
This vector potential for a uniform axial magnetic field, employing the symmetric gauge, 
results ${\mathbf A}=\frac{B}{2} (-y,x,0)$, with $B$ being the magnetic field strength. 
Finally, $U$ stands for the Hubbard repulsion parameter.
 
Obviously, the system dynamics does not depend on the selected gauge for the potential vector and the selected origin of coordinates. 
We thus set the origin of coordinates at the center of the cyclic molecule in our three-terminal devices.
Since the molecule can be described by a regular polygon, 
it is straightforward to show that all nearest-neighbor $t_{ij}$ tunneling parameters connecting two 
neighbor polygon vertices are the same, once the index rotation sense is fixed. 
Note that $t_{ji}=t_{ij}^*$, and again, the  dynamics of the system do not depend on the selected sense of rotation. 
Further, since the Hubbard Hamiltonian (\ref{eq1}) only depends on the topology, 
one may assume --without loss of generality-- that the input and output channels (or leads) are arranged radially 
from the vertices of the polygon. 
Because the in-plane coordinates $x$ and $y$ along a radial line are related by $y=a x$,
the Peierls phase between consecutive sites $i$ and $j$ of such line becomes zero,
as $\int_i^j {\mathbf A} \cdot d{\mathbf l}= \frac{B}{2} \int_i^j (-a x,x,0) \cdot (dx,a dx,0)=0$. 
In other words, the tunneling parameter between two lead sites or between a polygon site and a neighbor lead site can be set to $t^0_{ij}$,
regardless of the field and the actual geometry.

To solve Hamiltonian (\ref{eq1}) for $N$ electrons, the Hamiltonian is expanded onto the 
complete full configuration interaction (FCI) space containing 
$\Omega=\left(\begin{array}{c} 2K\\N\end{array}\right)$ Slater determinants 
$D_i(1,2,\dots, N)=|\chi_i(1)\sigma_i(1)* \chi_j(2)\sigma_j(2) * \dots * \chi_k(N)\sigma_k(N)|$, 
where $K$ is the number of independent particle functions. 
$K$ is also the number of sites in the system, as our TB model considers a single independent 
particle orbital $\chi_i$ centered at the site $i$. The wave function is then:
\begin{equation}
\label{eq2}
\Psi(1,2,\dots ,N)=\sum_i^{\Omega} c_i(t) \, D_i(1,2,\dots, N)
\end{equation}
The expectation value of the density operator $\hat \rho({\mathbf r})=\sum_i^N \delta({\mathbf r}-{\mathbf r}_i)$ results:
\begin{equation}
\label{eq3}
\langle \Psi| \hat \rho|\Psi\rangle=\sum_{ij}^N c_i(t)^* c_j(t) \langle D_i|\hat \rho({\mathbf r})|D_j \rangle
\end{equation}
\noindent and the population of the site $a$ is 
\begin{equation}
\label{eq4}
\langle \Psi| \hat \rho|\Psi\rangle_a=\sum_{i}^N |c_i(t)|^2(\delta_{a,\alpha_i}+\delta_{a,\beta_i})
\end{equation}
\noindent where  $\delta_{a,\alpha_i}$ ($\delta_{a,\beta_i}$) is zero unless $\chi_a$ is present in $D_i$ with spin $\alpha$ ($\beta$).\\

The time-dependent Schr\"odinger equation for Hamiltonian (\ref{eq1}) projected on the wave function (\ref{eq2}) can be written 
in atomic units as
\begin{equation}
\label{eq5}
i \, \frac{d}{dt} {\mathbb C}  ={\mathbb H} {\mathbb C},
\end{equation}
\noindent which is equivalent to:
\begin{equation}
\label{eq6}
{\mathbb C}(t)  ={\mathbb U}(t) {\mathbb C}(0)
\end{equation}
\noindent where ${\mathbb U}(t) = \exp(-i \, t \, {\mathbb H})$ is the time evolution operator.
Note that Eq.~(\ref{eq6}) is equivalent to Eq.~(\ref{eq5}) because we do not introduce time-dependent parameters in the Hamiltonian.\cite{foot1}

It is convenient to define effective tunneling and time scales as follows. 
We factorize the tunneling parameter $t^0$ in the Hamiltonian: $H=t^0 H_0$, 
where $H_0$ is now the Hamiltonian $H$ in effective units (e.u.) 
Next, we replace $t*t^0 \to t $ in ${\mathbb U}(t)$, so that time is now given in e.u. 
Thus, we melt the dependence of the time-evolution operator in a single time-like parameter $t$ (e.u.) 
that makes changes in the system within the length scale, as we will see in section \ref{s:res}.

\subsection{Alternance symmetry}

The alternance symmetry is closely related to the topological properties of the TB approximation. 
This symmetry is related to the possibility to divide centers into two disjoints sets $C^\star$ and $C^\circ$ 
in such a way that any center of one set (say $C^\star$) can only tunnel to centers of the other set ($C^\circ$) and vice versa, 
so that the complete system has a bipartite graph structure.  
Although strictly speaking this symmetry --and the ensuing invariance properties-- is only be present when the system 
is described by an approximate model Hamiltonian, it may prove to be of great value in the classification of corresponding eigenstates, 
and the resulting selection rules in an interpretation of various spectral characteristics of the system. 
In other words, it has an indisputable physical origin. The importance of the approximate selection rules which result from this 
symmetry has been long recognized in one-photon absorption and emission spectroscopy,\cite{PariserJCP} 
magnetic circular dichroism,\cite{MichlJCP} two-photon absorption spectroscopy\cite{CallisJPC} etc. 

Alterance symmetry was first studied in the independent particle H\"uckel Hamiltonians,\cite{coulson} 
and later for the interacting particles Pariser-Parr-Pople (PPP) Hamiltonian,\cite{PariserJCP,pople} 
these papers being complemented by others in a proper assessment of the meaning, scope, connexions and 
relevance of this symmetry (see e.g. Refs.~\onlinecite{paldus1,paldus2}). 
The Hubbard Hamiltonian\cite{hubbard} can be considered an approximate PPP Hamiltonian where some minor 
terms in the repulsion part of the Hamiltonian have been neglected. 
In turn, TB can be considered an approximate PPP Hamiltonian where all repulsion terms are neglected. 
Since the alternance symmetry is related to the topology, then all three Hamiltonians display the symmetry. 

Alternance symmetry implies one-electron pairing properties. Namely, the independent particle eigenfunctions 
$|a_i\rangle$ and their corresponding energies  $\varepsilon(|a_i\rangle)$ in an alternant system are paired 
in such a way that for each eigenfunction  $|a_i\rangle= \sum_{\mu} c_{i \mu}|\mu\rangle$, 
where  $|\mu\rangle$ are the one-site TB functions and $c_{i \mu}$ is the coefficient of the expansion of
$|a_i\rangle$ in terms of the basis set $\{|\mu\rangle\}$, we can associate the alternant conjugate 
eigenfunction $|\tilde{a}_i\rangle= \sum_{\mu} \tilde{c}_{i \mu}|\mu\rangle$ such that 
$\tilde{c}_{i \mu}=c_{i \mu}$ if $\mu \in C^\circ$ and $\tilde{c}_{i \mu}=-c_{i \mu}$ if $\mu \in C^\star$. 
In addition, $\varepsilon(|\tilde{a}_i\rangle)=-\varepsilon(|a_i\rangle)+\rm{constant}$. 
Interestingly, since this symmetry is related to the topological properties of the TB it can be modulated 
by the magnetic field, because it comes into the Hamiltonian as a phase in the tunneling integral.

\section{Results}
\label{s:res}

\subsection{Single-electron transport}

\begin{figure}[h]
\includegraphics[width=0.45\textwidth]{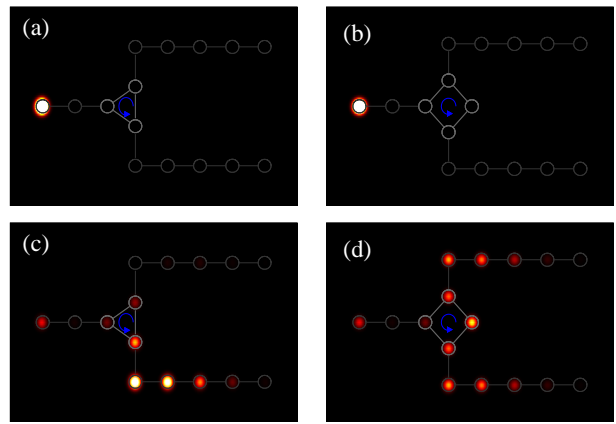}
\caption{Electron density at different times in three-terminal devices
under a magnetic flux of $\phi=0.25\phi_0$.
(a) and (c): triangular system at $t=0$ and $t=3$ e.u., respectively.
The magnetic flux favors transport through the lower output channel.
(b) and (d): rhombic system at $t=0$ and $t=3$ e.u., respectively.
The transport is symmetric inspite of the magnetic flux.
}\label{fig2}
\end{figure}

We start by investigating electron transport across a triangular cycle.
This is the simplest non-alternant system, its behavior being representative
of that found in other non-alternant cycles.
A perpendicular magnetic field is applied which gives rise to a tunneling parameter 
$t^0_{ij}=t^0 \, e^{i\, \frac{2 \, \pi}{3}  \frac{\Phi}{\Phi_0} }$,
where $i$ and $j$ are any two nearest-neighbor sites of the triangle.
The cyclic system is attached to one input and two output channels
arranged symmetrically, with one electron initially prepared in the
first site of the incoming channel, as schematically depicted in Fig.~\ref{fig2}(a).
For the actual calculations, the output channels have ten sites each.
We assume all the sites have the same potential, $\epsilon_i=\epsilon=0$,
and the hopping parameter is $t^0=-1$ e.u.

 With evolving time, the electron travels along the input channel and 
reaches the triangular system. At this point, the electron feels the
magnetic flux favoring a given sense of rotation (represented by the
blue curved arrow in the figure). For $\phi=0.25\,\phi_0$, the flux strongly
favors electron transfer into the lower channel. This can be seen in
Fig.~\ref{fig2}(c), which shows the electron density at a finite time 
when the electron leaves the cycle. If a flux of $\phi=-0.25\,\phi_0$ was
used, the transport would take place through the upper channel instead.

For comparison, we next investigate electron transport across a rhombic cycle.
Again, this is the simplest alternant system and it is representative of 
more complex alternant cycles. The tunneling parameter is now
$t^0_{ij}=t^0 \, e^{i\, \frac{2\pi}{4}  \frac{\Phi}{\Phi_0} }$.
Figure \ref{fig2}(b) shows the initial setup and Fig.~\ref{fig2}(d)
the corresponding electron density at a finite time.
Clearly, the electron transport is in sharp contrast with that of the
triangle, the transfer probability through upper and lower
channels now being identical.

Three-terminal devices have been employed by several groups both in mesoscopic and 
molecular systems (for reviews see e.g. Refs.~\onlinecite{TaoNNT,HodJPCM}),
and the possibility to switch the direction of electron transport magnetically
in such systems had been predicted by Hod et al., who further dicussed the 
convenience of the magnetic modulation as an alternative to electrical 
manipulation.\cite{HodJACS,HodJPCM}
In this context, Fig.~\ref{fig2} reveals that the magnetic control of the 
electron transport can only be achieved in non-alternant systems, 
thus establishing a critical parameter in the design of molecular junctions. 
This is the central finding of this work. 
In what follows we will gain deeper understanding on this
phenomenon and discuss its implications.

\begin{figure}[h]
\includegraphics[width=0.45\textwidth]{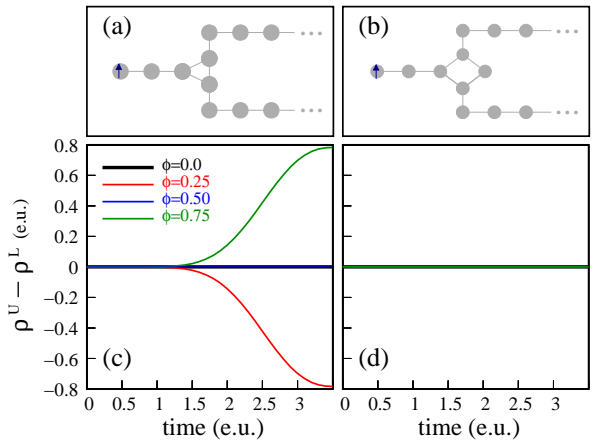}
\caption{(a-b) schematic of the systems under study at $t=0$, and (c-d) 
corresponding differences between electron density in the upper and lower 
channel as a function of time for different magnetic fluxes. The fluxes
are in flux quantum ($\phi_0$) units. 
}\label{fig3}
\end{figure}

In order to generalize the result of Fig.~\ref{fig2}, we next compare electron
transport in the triangular and rhombic systems for several values of the magnetix 
flux $\phi$. The systems under study are represented schematically in Figs.~\ref{fig3}(a-b), 
which show the setups at the initial time, $t=0$. To monitor the dynamics, we compute
the electron density in the upper and lower channels as 
$\rho_i= \sum_a \langle \Psi| \hat \rho|\Psi\rangle_a$,
where $i=U,L$ denotes the channel and $a$ runs over all the sites of the channel.
Note that for long times, when the charge density has left the cyclic system,
$\rho_i$ can be identified with the transfer probability through the $i$-th channel, $T_i$.
The conductance is proportional to this magnitude.\cite{ButtikerPRA}

Figure \ref{fig3}(c) shows the difference between the density in the upper and lower
channels as a function of time for the triangular cycle. In the absence of magnetic
flux, $\phi=0\phi_0$, the transport is symmetric through both output channels.
Switching on a positive magnetic flux $\phi=0.25\phi_0$, most of the transport takes place
along the lower channel, as anticipated in Fig.~\ref{fig2}(c). However, for $\phi=0.5\phi_0$
the transport becomes again symmetric, and for $\phi=0.75\phi_0$ the behavior reverses,
with transport taking place mostly along the upper channel. 

The influence on the magnetic flux on the sense of electron rotation can be understood
easily by considering the triangle as a closed system.
In the basis of atomic sites and using atomic units, the resulting eigenstates are:
\begin{equation}
\Psi_m = \left( 
\begin{array}{c}
e^{i(m+\phi)\,0} \\ 
e^{i(m+\phi)\,2\pi/3} \\ 
e^{-i(m+\phi)\,2\pi/3} \\ 
\end{array}
\right),
\end{equation}
\noindent associated to energies $\varepsilon_m = \varepsilon + 2\,t^0\, \cos{\left(\frac{2\pi (m+\phi)}{3}\right)}$, 
with $m=0\,\pm 1$. Although in triangles (finite-edges polygons in general) the angular momentum $L_z$ is not
a constant of motion, we may still relate $m>0$ ($m<0$) to anti-clockwise (clockwise) rotation.\cite{ring_limit}
Therefore, when $\phi=0$, $\Psi_1$ and $\Psi_{-1}$ become degenerate, so that this energy level does 
not have a preferred sense of rotation. The same happens to the non-degenerated $m=0$ level.
This is the underlying physical reason for the symmetric evolution on time of an state in which the electron is initially
set at one site of the triangle that evolves symmetrically to the other two sites.

Switching on the magnetic field introduces an effective angular momentum $m_\phi=(m+\phi)$
and lifts degeneracies, resulting in non-zero angular momenta and a favored 
direction of rotation. A special situation is found at half-integer values of the flux. 
Then, degeneracy between states with opposite effective angular momenta is found
(e.g. $m_\phi=0.5$ and $m_\phi=-0.5$ at $\phi=0.5$), and again there is no net angular
momentum. As a consequence, transport is again symmetric in both senses.
In short, the direction of electron transport in the triangle is determined by the 
combined effect of angular momenta and magnetic flux. It is symmetric 
when $\phi=k/2$ with $k$ integer, and asymmetric otherwise. 
Anti-clockwise (clockwise) rotation is favored for $k/2 < \phi < (k+1)/2$ 
when $k$ is even (odd), with a maximum at $\phi=k/4$.

We now investigate transport through a rhombic cycle, Fig.~\ref{fig3}(d).
The striking result is that, contrary to the case of the triangle, 
the dynamics is completely symmetric for any value of the magnetic flux.
In other words, using a cycle with an even number of sites removes the magnetic 
modulation of electron transport direction. This is a consequence of the alternance symmetry
and its relation to time-reversal symmetry, as we show next.\cite{preforgue}

Let us consider a TB alternant system, in the presence of a magnetic field. 
We set the origin of energies at the center of the energy spectrum so that $\epsilon_j=-\epsilon_{K-j}$, 
where $K$ is the number of sites in the system and hence also the dimension of the basis set. 
We can define the operator $\hat T$ acting on the independent particle functions 
$\chi_j$ of the system:
\begin{equation}
\label{eq_trev}
\hat T \; \chi_j \, e^{-i \epsilon_j t}=  \chi_{K-j} \, e^{i \epsilon_j t}=\chi_{K-j} \, e^{-i \epsilon_{K-j} t} 
\end{equation}
\noindent This operator does not commute with the Hamiltonian, but rather anticommutes with it, $\{\hat T, H\}=\hat T H+H \hat T=0$. 
Then, it represents a symmetry of the time-dependent Schr\"odinger equation, which transforms 
$i \, d/dt \, \Psi_j = H \Psi_j$ into $i \, d/dt \, \Psi_{K-j} = H \Psi_{K-j}$. 
This allows us to define a partition of the wave function space into a symmetric $\Psi_+=\Psi_j+\Psi_{K-j}$ 
and an antisymmetric $\Psi_-=\Psi_j-\Psi_{K-j}$ part. 
Since $\Psi_j=\sum_i^{\{C^\circ\}}a_i \; \chi_i^{\circ}+\sum_j^{\{C^\star\}}b_j \; \chi_j^{\star}$ and  
$\Psi_{K-j}=\sum_i^{\{C^\circ\}}a_i \; \chi_i^{\circ}-\sum_j^{\{C^\star\}}b_j \; \chi_j^{\star}$, 
the functions $\Psi_+/\Psi_-$ have only non-zero coefficients at the sites $C^\circ/C^\star$. 
Thus, preparing the electron initially in a $C^\circ$ ($C^\star$) site (or several sites of the same kind) 
means the system is in a state defined by $\Psi_+(0)$ ($\Psi_-(0)$).
Because $ \epsilon_j = -\epsilon_{K-j}$, the system will experience, as time runs, 
a synchronic decrease of population in the sites $C^\circ$ ($C^\star$) and a simultaneous increase of 
it at the partner sites $C^\star$ ($C^\circ$). 
Note that this will happen irrespectively that an axial magnetic field, favoring a given sense of electronic circulation, 
acts or not on the system.

\subsection{Few-electron transport}

\begin{figure}[h]
\includegraphics[width=0.45\textwidth]{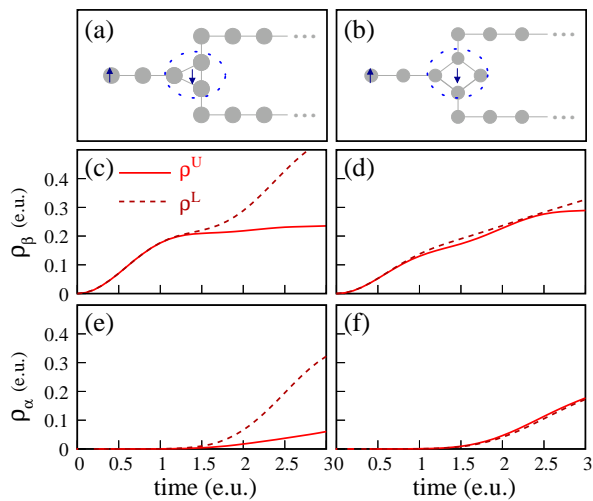}
\caption{(a-b): schematic of the systems under study at $t=0$. 
(c-d): corresponding differences between spin $\beta$ 
electron density in the upper and lower channel as a function of time.
(e-f): same but for spin $\alpha$ electron.
In all cases $\phi=0.25\phi_0$.
Note that $\rho_\alpha$ monitors the electron initially localized in the
first site of the incoming channel, while $\rho_\beta$ does so for the
electron initially delocalized over the cylic system. 
}\label{fig4}
\end{figure}

We next investigate the effect of electron-electron repulsion on the
magnetotransport of our three-terminal devices.
To this end we add a resident electron delocalized over the cyclic system,
as plotted in Figs.~\ref{fig4}(a-b).
The Hubbard repulsion parameter is set to $U=10\,t^0$, the ratio $U/|t^0|=10$ thus being of the same order 
of magnitude as that employed in molecular systems\cite{parr} and quantum dots.\cite{DelgadoPRL} 

The resident electron has opposite spin ($\beta$) to that of the incoming electron ($\alpha$).
This allows us to track their time evolutions separately by plotting the spin polarized 
electron densities. Thus, for the triangular system, Fig.~\ref{fig4}(c) shows the 
time evolution of the resident electron (spin $\beta$ electron density), while 
Fig.~\ref{fig4}(e) does so for the incoming one (spin $\alpha$ density).
We have set $\phi=0.25\,\phi_0$. 
One can see that the resident electron spreads into the output channels symmetrically
for times $t<1.5$ e.u. This is in spite of the system being non-alternant.
The reason is that the electron is initially delocalized in all three sites of the
triangle, so that the magnetic field induced rotation favors transport equally through
all three channels, much as in a sprinkler.
For longer times, however, transport through the lower channel is suddenly favored.
It is worth noting that this coincides with the arrival of the spin $\alpha$ electron,
see panel (e). This is because the spin $\beta$ density first scattered across the input
channel, bounces back when it meets the spin $\alpha$ electron owing to Coulomb repulsion.
It then reenters the system core and travels as in the single-electron case described in
the previous subsection, i.e. mostly through the lower channel. Also the incoming $\alpha$
electron follows this path.

For the rhombic system, Fig.~\ref{fig4}(f) shows the incoming electron travels
symmetrically, as in the single-electron system. Instead, Fig.~\ref{fig4}(d) shows
that the resident electron transport is no longer exactly symmetric. This is
because the resident electron initially occupies both $C^\circ$ and $C^\star$ sites,
so its wave function cannot be described by $\Psi_+$ or $\Psi_-$ alone.
In other words, alternance symmetry is broken, and hence symmetric transport is no
longer granted.

\begin{figure}[h]
\includegraphics[width=0.45\textwidth]{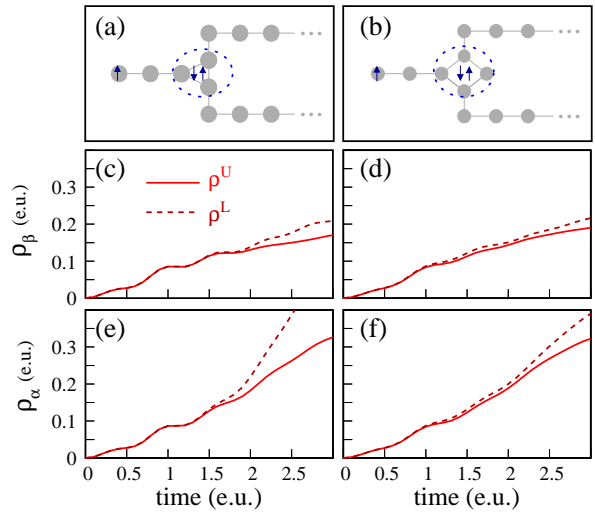}
\caption{Same as Fig.~\ref{fig4} but with two resident electrons forming a singlet.}\label{fig5}
\end{figure}

Similar findings are obtained if the system contains two spin-paired electrons.
Fig.~\ref{fig5} shows the corresponding results. For short times ($t<1.5$ e.u.),
transport is symmetric through both output channels owing to the sprinkler-like effect. 
By comparing $\rho_\alpha$ and $\rho_\beta$, no spin polarized transport is observed
in this regime. For longer times, when the incoming electron arrives, the transfer 
probability increases in the lower lead.

We conclude from this subsection that the behavior of interacting electrons 
essentially follows an independent particle scheme, except for the presence of 
the Coulomb hole. The reason is that for a ratio $U/|t^0|=10$ double-occupancy states are
at much higher energy than single-occupancy ones. Then, the consevation of the average energy
$\langle E \rangle$ along time only allows minute contributions of these states in $\Psi(t)$,
so the system evolution resembles that of independent particles.
In setups with lower $U/|t^0|$ ratio, Coulomb correlations may play a role,
but in the limit of low $U/|t^0|$ ratio, electron-electron interactions become
negligible and we are again in the independent particle scheme.
Thus, in most cases the role of alternance symmetry on the electron transport is 
captured by the single-electron picture.

\section{Discussion}
\label{s:disc}

In this section we briefly discuss some implications of alternance symmetry on
the electron magnetotransport through existing or potentially interesting molecular 
and mesoscopic systems.

\subsection{Molecular junctions}

As mentioned earlier, while the role of spatial symmetry in molecular junctions is 
under active research,\cite{MarkussenNL,TaniguchiJACS,LortscherCPC} 
that of alternance symmetry has been overlooked so far. 
As shown in the previous section, alternance symmetry is critical in determining
the magnetoconductance of cyclic systems. 
In particular, its absence (presence) enables (disables) magnetic control of the 
electron transport direction.
This kind of control has been proposed as an alternative to electric control.\cite{HodJPCM}
In the case of molecules, the main handicap is that very strong magnetic fields are required to achieve
significant flux piercing the small area of usual molecules, such as benzene.
This problem can be overcome by using macromolecules with larger areas.
Several systems are suited to this end, including quantum corrals made of metal atoms,\cite{NazinSCI}
polyaromatic hydrocarbons\cite{MardaaniJMMM} colloidal graphene quantum dots,\cite{MuellerNL}
or nanographene rings.
As a matter of fact, it has been recently recognized that the latter structures
often present defects in the form odd-membered rings.\cite{KawasumiNC}
This clearly breaks the alternance symmetry, and the ensuing consequences on 
the transport should be born in mind.

\begin{figure}[h]
\includegraphics[width=0.45\textwidth]{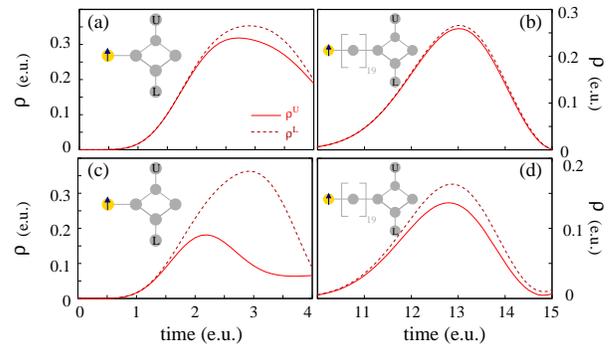}
\caption{Electron density in the upper and lower channel as a function of
time at $\phi=0.25\phi_0$. (a) and (b): the injected electron is in a heteroatom
with $\Delta \epsilon=0.1\,t^0$. 
(c) and (d): the heteroatom has $\Delta \epsilon = t^0$.
The insets show the structure under study. Left (right) column: 
the heteroatom is next to (20 sites away from) to the cycle. 
}\label{fig6}
\end{figure}

Hod and co-workers have proposed a magnetoresistance logic gate based on a 
three-terminal device containing a macrocycle composed of 48 benzene rings.\cite{HodJACS}
The operating principle relied on the asymmetric transport through the two
output leads induced by an external magnetic field. 
It is worth noting that they expect asymmetric transport inspite of having 
considered an even-membered ring. The reason is that the the macromolecule
is attached to gold leads. In our work we have assumed that all sites have
the same potential, $\epsilon$. Introducing sites with different energy,
such as carbon and gold atoms, renders the energy spectrum asymmetric
with respect to its center, hence breaking the symmetric transport.
This effect is strong when the heteroatoms are in contact with the cycle
and gradually fades when away from it, as can be seen in Fig.~\ref{fig6}
for the systems in the insets.
The figure shows the electron density reaching the upper and lower channels
when the electron is initially injected in a heteroatom which is one site
(left panels) and twenty sites (right panels) away from the alternant cycle.
Typical heteroatoms in conjugated molecules have energies which differ
from those of carbon atoms by $\Delta \epsilon \approx 0.5 - 2\,t^0$.\cite{streitwieser_book} 
We then consider two possible orders of magnitude, $\Delta \epsilon = 0.1\,t^0$
(top panels) and $\Delta \epsilon = 1.0\,t^0$ (bottom panels).
When the heteroatom is next to the cycle (panels (a) and (c)), one observes
asymmetry not only in the amount of density reaching the upper and lower channels,
but also in the time this occurs. Clearly, the asymmetry is more pronounced in panel (c),
indicating that the nature of the heteroatom (and hence $\Delta \epsilon$) is critical in determining the extent of the asymmetry. 
Still, when the heteroatom is away from the cycle (panels (b) and (d)) the asymmetry
is visibly reduced. 
This suggests that, even in the presence of heteroatoms, nearly symmetric transport
associated with alternance symmetry can be expected if the input channel is long enough.

As opposed to the potential of the sites $\epsilon$, varying the tunneling parameter 
$t^0_{ij}$ of the leads with respect to that of the cycle does not induce any asymmetric transport.

As an illustration of the applications of alternance symmetry, with the above 
considerations, one could envisage a molecular device with selective electron
transport for photocatalysis or photocurrent generation. Electrons are generated in
a photoactive group at the far end of the input chain. Next they
travel until a cyclic macromolecule subject to a constant magnetic field. 
The molecule has two additional (output) chains contacting the electrodes.
If the molecule is non-alternant, the magnetic field will drive electron
transport preferentially through one of the chains. If it is alternant, 
the transport will be symmetric instead, although one can easily switch to
non-alternant e.g. by chemical substitution of the aromatic cycles.

\subsection{Coupled quantum dots}

Contrary to molecular systems, the magnetoconductance of coupled quantum dot 
systems has been thoroughly investigated theoretically and experimentally. 
Several studies have dealt with triangular triple-dot structures,\cite{GaudreauPRB,DelgadoPRL,JiangJPCM,EmaryPRB,KuzmenkoPRL}
but squared four-dot cycles have been also proposed.\cite{ScarolaPRL,ZengPRB}
Much of the interest in these systems relied on the influence of the 
AB effect on the electron transport.

The large magnetic fluxes that can be achieved in these systems
make them ideal to experimentally test the role of alternance symmetry
described in this work. The main difficulty will be to produce truly 
alternant systems. The chemical potential of the leads can be
set to match that of the quantum dots, so that all sites have
the same energy, while electrostatic gates can be used to minimize
the energy difference between the dots, which are otherwise different
owing to size and composition inhomogeneities. Yet, electrostatic
fluctuations may play a role.\cite{KlimeckPRB}

We close by noting that the magnetotransport we report here 
is a pure quantum mechanical effect, derived from the non-trivial AB 
phase factor introduced by the magnetic flux piercing the cyclic system. 
In finite-width quantum dots, the magnetic field also pierces the dot
itself and an additional asymmetry in the transport may arise from
the Lorentz force acting on the electrons, as noticed in quantum rings.\cite{SzafranEPL}

\section{Conclusions}
\label{s:conc}

In conclusion, using a Hubbard model we have shown that the sense
of electron transport across a cyclic system can be governed by 
magnetic fields only if it is non-alternant.
The presence of alternance symmetry imposes symmetric transport
in all directions.
This result is independent of the magnetic field value and it holds 
both for single-electron and few-interacting electron systems.
We then argue that this topological symmetry is a critical parameter
in the design of atomic or mesoscopic molecular junctions for 
magnetoconductance devices.

\begin{acknowledgments}
Support from MICINN project CTQ2011-27324 and UJI-Bancaixa project P1-1B2011-01
is acknowledged.
\end{acknowledgments}


\begin{thebibliography}{50}

\bibitem{GaudreauPRB}
L. Gaudreau, A. S. Sachrajda, S. Studenikin, A. Kam, F. Delgado, Y. P. Shim, M. Korkusinski, and P. Hawrylak,
Phys. Rev. B {\bf 80}, 075415 (2009). 

\bibitem{TarankoJAP}
E. Taranko, M. Wiertel, and R. Taranko,
J. Appl. Phys. {\bf 111}, 023711 (2012) and references therein.

\bibitem{GaudreauNP}
L. Gaudreau, G. Granger, A. Kam, G. C. Aers, S. A. Studenikin, P. Zawadzki, M. Pioro-Ladriere, Z. R. Wasilewski, A. S. Sachrajda, 
Nature Phys. {\bf 8}, 54 (2012).

\bibitem{LidarPRL}
D. A. Lidar, I. L. Chuang, and K. B. Whaley, 
Phys. Rev. Lett. {\bf 81}, 2594 (1998).

\bibitem{ScarolaPRL}
V. W. Scarola, K. Park, and S. Das Sarma,
Phys. Rev. Lett. {\bf 93}, 120503 (2004).

\bibitem{NitzanSCI}
A. Nitzan, and M. A. Ratner,
Science {\bf 300}, 1384 (2003).

\bibitem{TaoNNT}
N. J. Tao, 
Nature Nanotechnol. {\bf 1}, 173 (2006).

\bibitem{SongAM}
H. Song, M. A. Reed, and T. Lee,
Adv. Mater. {\bf 23}, 1583 (2011).


\bibitem{LiuJACS}
C. Liu, D. Walter, D. Neuhauser, and R. Baer,
J. Am. Chem. Soc. {\bf 125}, 13936 (2003).

\bibitem{HodJACS}
O. Hod, R. Baer, and E. Rabani,
J. Am. Chem. Soc. {\bf 127}, 1648 (2005).

\bibitem{HodJPCM}
O. Hod, R. Baer, and E. Rabani,
J. Phys.: Condens. Matter {\bf 20}, 383201 (2008).

\bibitem{HansenJCP}
T. Hansen, G. C. Solomon, D. Q. Andrews, and M. A. Ratner,
J. Chem. Phys. {\bf 131}, 194704 (2009).

\bibitem{MarkussenNL}
T. Markussen, R. Stadler, and K. S. Thygesen,
Nano Lett. {\bf 10}, 4260 (2010).

\bibitem{AharonovPR}
Y. Aharonov and D. Bohm,
Phys. Rev. {\bf 115}, 485 (1959).

\bibitem{MardaaniJMMM}
M. Mardaani, and H. Rabani,
J. Magn. Magn. Mater. {\bf 331}, 28 (2013).

\bibitem{DelgadoPRL}
F. Delgado, Y. P. Shim, M. Korkusinski, L. Gaudreau, S. A. Studenikin, A. S. Sachrajda, and P. Hawrylak,
Phys. Rev. Lett. {\bf 101}, 226810 (2008).

\bibitem{JiangJPCM}
Z. T. Jiang, Q. F. Sun,
J. Phys.: Condens. Mater. {\bf 19}, 156213 (2007).

\bibitem{EmaryPRB}
C. Emary,
Phys. Rev. B {\bf 76}, 245319 (2007).

\bibitem{KuzmenkoPRL}
T. Kuzmenko, K. Kikoin, and Y. Avishai,
Phys. Rev. Lett. {\bf 96}, 046601 (2006).

\bibitem{WebbPRLTimpPRL}
R. A. Webb, S. Washburn, C. P. Umbach, and R. B. Laibowitz,
Phys. Rev. Lett. {\bf 54}, 2696 (1985);
G. Timp, A. M. Chang, J. E. Cunningham, T. Y. Chang,
P. Mankiewich, R. Behringer and R. E. Howard 
Phys.  Rev. Lett. {\bf 58} 2814 (1987).

\bibitem{ChwiejJPCMIjasPRB}
T. Chwiej, and B. Szafran,
J. Phys.: Condens. Matter {\bf 25}, 155802 (2013);
M. Ij\"as, A. Harju, 
Phys. Rev. B {\bf 85}, 235120 (2012)

\bibitem{SzafranEPL}
B. Szafran, and F. M. Peeters,
Europhys. Lett. {\bf 70}, 810 (2005).

\bibitem{TaniguchiJACS}
M. Taniguchi, M. Tsutsui, R. Mogi, T. Sugawara, Y. Tsuji, K. Yoshizawa, T. Kawai,
J. Am. Chem. Soc. {\bf 133}, 11426 (2011).

\bibitem{LortscherCPC}
E. L\"ortscher,
Chem. Phys. Chem. {\bf 12}, 2887 (2011).

\bibitem{Orchin_book}
M. Orchin, R.S. Macomber, A.R. Pinhas, and R.M. Wilson, 
\emph{The Vocabulary And Concepts of Organic Chemistry} (John Wiley \& Sons, New Jersey, 2005).

\bibitem{PariserJCP} 
R.Pariser,
J. Chem. Phys. {\bf 24}, 250 (1956).

\bibitem{MichlJCP} 
J. Michl, 
J. Chem. Phys. {\bf 61}, 4270(1974); 
R.J. van der Wal and P. J. Zandstra,
ibid. {\bf 64}, 2261 (1976).

\bibitem{CallisJPC}
P. R. Callis, T. W. Scott, and A. C. Albrecht, 
J. Phys. Chem. {\bf 78}, 16 (1983);
R. P. Ravaand, and L. Goodman,
J. Am. Chem. Soc. {\bf 104}, 3815 (1982); 
B. Dick and G. Hohlneicher, 
Chem. Phys. Lett. {\bf 97}, 324 (1983).

\bibitem{Jones_book}
R. A. Y. Jones, \emph{Physical and Mechanistic Organic Chemistry}, 
(Cambridge Univ. Press, Cambridge, 1979).

\bibitem{GrafPRB} 
M. Graf and P. Vogl,
Phys. Rev B {\bf 51}, 4940 (1995).

\bibitem{foot1} 
Otherwise one should use $U(t)=\Pi_j U(t_j+\Delta t,t_j)$, with $\Delta t$ small enough to ensure a correct chronological order of the infinitessimal time-evolution operators. Alternatively, one could employ the Magnus expansion of the time evolution operator (see e.g. Ref.~\onlinecite{tannor_book})

\bibitem{tannor_book} 
D. J. Tannor,
{\it Introduction to Quantum Mechanics: A Time-Dependent Perspective},
(University Science Books, Sausalito, 2006).

\bibitem{coulson} 
C. A. Coulson and G. S. Rushbrooke, 
Proc. Camb. Phil. Soc. {\bf 36} 193 (1940).

\bibitem{pople} 
J. A. Pople,
Trans. Faraday Soc. {\bf 49} 1375 (1953).

\bibitem{paldus1}
J. \u{C}\'{\i}\v{z}ek, J. Paldus and I. Huba\u{c},
Int. J. Quantum Chem. {\bf 8},951 (1974). 

\bibitem{paldus2}
J. Kouteck\'y and J. Paldus,
J. Chem. Phys. {\bf 83} 1722 (1985). 

\bibitem{hubbard}
J. Hubbard, J.
Proc. R. Soc. Lond. A {\bf 276} 238 (1363).

\bibitem{ButtikerPRA}
M. B\"uttiker, Y. Imry, and M. Y. Azbel,
Phys. Rev. A {\bf 30}, 1982 (1984).

\bibitem{ring_limit}
For a $k$-sites-polygon $\Psi_m $ has $k$ components $e^{i(m+\phi)\, \theta}$, where 
$\theta=(2\pi/k) \, q$ with $q=0, \pm1, \pm2, \dots, \pm (k-1)/2$ for odd $k$ and $q=0, \pm1, \pm2, \dots \pm k/2$ 
 for even $k$. As $k$ rises up to infinity then, $\theta$ becomes a continuous variable and we end up with the 
eigenfunctions of a particle in a ring. 

\bibitem{preforgue}
A different, yet somehow equivalent, connexion between alternance and time-reversal symmetry is discussed in 
 A. Laforgue, J. \u{C}\'{\i}\v{z}ek and J. Paldus
J. Chem. Phys. {\bf 59} , 2560 (1973).

\bibitem{parr}
In molecules the one-site repulsion $U$ is generally estimated by means of the I-A approximation, yielding $U=10.84 \,eV$. The
hopping parameter employed $-2.29 \,eV$ is the so-called spectroscopic. See e.g. R. G. Parr, \emph{The Quantum Theory 
of Molecular Electronic Structure}, (Benjamin, New York, 1963).


\bibitem{NazinSCI}
G. V. Nazin, X. H. Qiu, and W. Ho 
Science {\bf 302} 77 (2003).

\bibitem{MuellerNL}
M. L. Mueller, X. Yan, J. A. McGuire, and L.S. Li,
Nano Lett. {\bf 10}, 2679 (2010).

\bibitem{KawasumiNC} 
K. Kawasumi, Q. Zhang, Y. Segawa, L. T. Scott, and K. Itami,
Nature Chem. {\bf 5}, 739 (2013).

\bibitem{streitwieser_book}
A. Streitwieser, \emph{Molecular Orbital Theory for Organic Chemists}, (Wiley, New York, 1961).

\bibitem{ZengPRB}
Z. Y. Zeng, F. Claro, and A. P\'erez,
Phys. Rev. B {\bf 65}, 085308 (2002).

\bibitem{KlimeckPRB}
G. Klimeck, G. Chen, and S. Datta,
Phys. Rev. B {\bf 50}, 2316 (1994).

\end{thebibliography}
\end{document}